\documentclass[prb,twocolumn,showpacs,preprintnumbers,amsmath,amssymb]{revtex4}
\usepackage{amsmath,amssymb,amsfonts}
\usepackage{bm}
\usepackage{graphicx}
\usepackage{times}
\usepackage{color}
\usepackage[colorlinks,bookmarks=false,citecolor=blue,linkcolor=red,urlcolor=blue]{hyperref}

\newcommand{\fref}[1]{Fig.~\ref{#1}}
\newcommand{\eref}[1]{Eq.~(\ref{#1})}

\newcommand{\normwidth}{0.8\columnwidth}

\newcommand{\figwidth}{0.97\columnwidth}

\definecolor{DarkGreen}{rgb}{0.0,0.7,0.0}
\definecolor{Grey}{rgb}{0.5,0.5,0.5}
\definecolor{Red}{rgb}{0.9,0.3,0.3}

\newcommand{\bbf}{\textrm}

\begin{document}

\title{Metallic surface of a bipolaronic insulator}
\author{Reza Nourafkan}
\affiliation{Department of Physics, University of Alberta, Edmonton, Alberta, Canada T6G 2G7}
\author{Frank Marsiglio}
\affiliation{Department of Physics, University of Alberta, Edmonton, Alberta, Canada T6G 2G7}
\author{Massimo Capone}
\affiliation{ISC-CNR, UoS Sapienza and Dipartimento di Fisica, Universit\`a Sapienza, P.le Aldo Moro 2, I-00185, Roma, Italy}

\begin{abstract}
We investigate the possibility that the surface of a strongly coupled
electron-phonon system behaves differently from the bulk when the relevant
parameters are  inhomogeneous due to the presence of the interface. We consider parameter variations which make the surface either more metallic or more insulating than the bulk. While it appears impossible to stabilize a truly insulating surface when the bulk is metallic, the opposite situation can be realized. A metallic surface can indeed be decoupled from a bipolaronic insulator realized in the bulk.
\end{abstract}
\maketitle

\section{Introduction}
The interest in the electronic properties of surfaces and interfaces
is growing due to the increasing ability to engineer interfaces
between correlated materials and to accurately measure surface and bulk properties. A number of discrepancies have been reported between bulk and surface properties of complex materials\cite{matzdorf,anisimov}, while interfaces between different materials can lead to surprising properties. A notable example is the metallic interface between the two insulators LaTiO$_3$ and SrTiO$_3$\cite{ltosto}.

On the theoretical side, the investigation of the effects of surfaces and interfaces has been focused on Hubbard-type
models, in which local repulsion correlates the electronic motion eventually leads, for commensurate densities, to a Mott insulating state when
the Coulomb interaction is sufficiently large. These studies have either used extensions\cite{Potthoff1,Ishida} of the dynamical mean-field theory (DMFT)\cite{Georges}, a theoretical approach which has provided the first unified scenario of the Mott transition, or variational approaches.\cite{Borghi} Studies of solid-vacuum interfaces have unveiled the possibility of surface ferromagnetism\cite{Potthoff2}, and have described the penetration depth of a bulk metallic phase into an otherwise insulating surface.\cite{Helmes} Indeed, Borghi \textit{et al}.\cite{Borghi} have shown the existence of a dead layer, due to an exponential penetration of metallic excitations.

Another localizing effect which affects the properties of electrons in solids is the electron-phonon (e-ph) interaction.
Also in this case quantum fluctuations inherent to the low dimensionality of surfaces and interfaces
and strong interactions can stabilize novel ground states that are distinct from the bulk. In Ref.\cite{matzdorf} the freezing of a bulk phonon at the surface has been invoked as the source of remarkable electronic properties.

Similarly to the case of repulsive electron-electron interactions, important insights into the problem of strongly
coupled e-ph
systems have been gained by DMFT\cite{Georges}. Studies of the
Holstein model in a homogeneous bulk system using DMFT\cite{CaponeCiuchi,Capone3,Holstein_altri} show that
as the e-ph interaction increases, the conduction electrons progressively lose their mobility, eventually evolving into a polaronic state
in which the presence of an electron is associated with a finite lattice distortion.
The same e-ph coupling can cause any two polarons to attract and form a bound pair in real space,
called a bipolaron. \bbf{When the number of carriers equals the number of
sites, i.e., the lattice is half-filled,} bipolaron formation causes the system to undergo a continuous (at zero temperature)
metal to insulator transition  at a critical e-ph coupling.

We have investigated the effect of a solid-vacuum interface on this
scenario in a previous paper, \cite{Nourafkan} in which, in order to
focus on the purely geometrical aspect of the problem we have
considered the same parameters in the surface and in the bulk.
Even for uniform parameters, the band narrowing at the surface\cite{Kalkstein} causes a reduction in quasiparticle weight relative to the bulk, i.e., the surface is less metallic. Therefore, upon increasing the e-ph interaction strength the polaron crossover takes place first on the surface layer. Nonetheless, for uniform model parameters, enhanced correlation effects at the surface are not sufficient to turn the surface insulating before the bulk (i.e., for a smaller e-ph coupling) and a single metal-insulator transition occurs at the critical coupling for the infinite system $g_c=g_{c,bulk}$.\cite{Nourafkan}

Besides the geometrical effect of missing neighbors, the surface properties are complicated by the fact
that the microscopic interactions close to the surface have a value
which may differ significantly from that in the bulk. In the Holstein
model a modification of the hopping as well as of the e-ph coupling
strength in the vicinity of the surface should be expected for any
real systems. The relaxation of the interlayer distance, for example,
can cause an enhancement or decrease in the hopping integrals at the
surface. In this work  we shall extend the analysis of
Ref. \cite{Nourafkan} to nonuniform model parameters and investigate
the possibility of the occurrence of a metallic surface concurrent
with a bulk bipolaronic insulator or of a bipolaronic insulating
surface concurrent with a normal metal in the bulk. We will consider
the half-filled case, in which an actual phase transition can be
observed, even if particle-hole symmetry forbids charge
transfer between surface and bulk.

The paper is organized as follows. In Sec. II we introduce the
model Hamiltonian, which is a semi-infinite Holstein model with layer dependent parameters. In addition, we briefly describe the embedding approach for DMFT. Results for a range of modified surface parameters are presented and discussed in Sec. III. In Sec. IV we summarize with some concluding remarks.

\section{Model and Method}
We investigate the Holstein model on a three-dimensional,
bipartite simple-cubic (sc) lattice with nearest-neighbor hopping. The lattice is cut along a plane perpendicular to one
of the coordinate axes, e.g., the $z$-axis [sc(001) surface].
The system is considered to be built up by two-dimensional layers parallel to the surface.
Accordingly, the position vector to a particular site in the
semi-infinite lattice is written as ${\bm R}_{site}={\bm r}_{i}+{\bm
R}_{\alpha}$. Here ${\bm R}_{\alpha}$ stands for the coordinate
origin in the layer $\alpha$ and the layer index runs from
$\alpha=1$ for the topmost surface layer to infinity. ${\bm r}_{i}$ is
the position vector with respect to a layer-dependent origin, and
runs over the sites within the layer. Each lattice site is then labeled by
indices $i$ and $\alpha$.
In this notation, the Hamiltonian reads:
\begin{eqnarray}
\label{eq:Holstein}
H=&-&\sum_{\langle i\alpha , j\beta\rangle \sigma} t_{i\alpha,j\beta} c^\dagger_{i\alpha \sigma}c_{j\beta \sigma} +\Omega_{0}\sum_{i\alpha} b^\dagger_{i\alpha}b_{i\alpha}\nonumber \\
&+& \sum_{i\alpha}g_{\alpha}{\left( n_{i\alpha} - 1\right)\left( b^\dagger_{i\alpha} + b_{i\alpha} \right)},
\end{eqnarray}
where $c_{i\alpha\sigma}\left(c^\dagger_{i\alpha\sigma} \right)$ and
$b_{i\alpha}\left(b^\dagger_{i\alpha} \right)$ are, respectively,
destruction (creation) operators for electrons with spin
$\sigma$ and local vibrons of frequency $\Omega_{0}$
on site $i$ of the $\alpha$ layer. The electron density on site $i\alpha$ is denoted $n_{i\alpha}$,
$t_{i\alpha,j\beta}$ is the hopping
matrix element between two nearest-neighbor sites, and $g_{\alpha}$
denotes the layer-dependent electron-phonon coupling strength. We
fix the energy scale by setting $t_{\langle
i\alpha,j\beta\rangle}\equiv t = 1$ for $\alpha,\beta \neq 1$.

\begin{figure}
\begin{center}
\includegraphics[width=\figwidth]{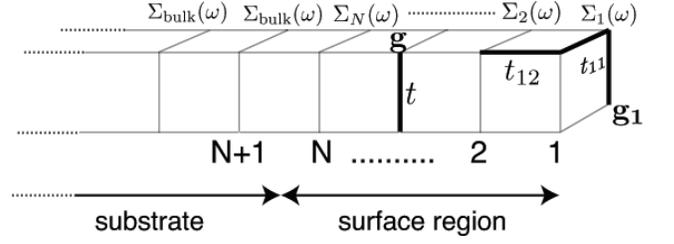}
\caption{Geometry of the (001) surface of a simple cubic lattice. The index $\alpha$ (horizontal axis) labels the layers parallel to surface. $\alpha=1$ refers to topmost layer. In the embedding approach for DMFT the system is divided into a surface region of $N$ layers and a semi-infinite substrate. The physical properties in the substrate (e.g., electron self-energy) are not layer-dependent and coincide with those of the bulk.}
\label{fig:fig0}
\end{center}
\end{figure}

To solve our model, we use an extension of DMFT to inhomogeneous systems called the embedding approach for DMFT.\cite{Ishida} In this scheme, the layered structure is partitioned into a surface region which includes the first $N$ layers, and the adjacent semi-infinite bulk region (substrate) which is coupled to it (see \fref{fig:fig0}). The surface corresponds to the region where one expects different properties relative to the bulk. It is shown next that the influence of the semi-infinite substrate on the surface region can be described in terms of an energy-dependent embedding potential. This can be viewed as an additional self-energy due to the transitions between the surface and the substrate.
Because of translational symmetry in the plane parallel to the
interface, the embedding potential of the substrate is diagonal with
respect to the two-dimensional wave vector ${\bf k}=(k_x,k_y)$ and
can be expressed as an $N\times N$ matrix.

By defining ${\bm A}({\bf k}, i\omega_n)=\big[(i\omega_n+\mu){\bm 1}-{\bm\epsilon}({\bf k})-{\bf \Sigma}(i\omega_n) \big]$, the equation for the Green's function is given by:
\begin{equation}
{\bm A}({\bf k}, i\omega_n){\bf G}({\bf k}, i\omega_n)= {\bm 1}.
\label{eq:green1}
\end{equation}
While the surface region consists of only $N$ layers, the matrices corresponding to the Green's function
are infinite dimensional due to the semi-infinite substrate.
In \eref{eq:green1}, ${\bf \Sigma}(i\omega_n)$ is the self-energy matrix, which in
the framework of single-site DMFT, is local
[i.e., $\Sigma(i\omega_n)_{\alpha\beta}=\Sigma_{\alpha}(i\omega_n)\delta_{\alpha\beta}$] and independent of wave vectors,
$\bf k$. The chemical
potential is given by $\mu$ and ${\bm \epsilon}({\bf k})$ is the two-dimensional
dispersion relation, which includes information about the surface
geometry. The ${\bm \epsilon}({\bf k})$ matrix for a surface cutting a simple cubic lattice
with a plane perpendicular to the $z$ direction [sc(001) surface] assumes the following form:\cite{Potthoff1}
\begin{eqnarray}
{\bm \epsilon}({\bf k})=\left(
\begin{array}{cccc}
t_{11}\epsilon_{\parallel}({\bf k}) & t_{12}\epsilon_{\perp}({\bf k}) & 0 &0 \\
t_{21}\epsilon_{\perp}({\bf k}) & t_{22}\epsilon_{\parallel}({\bf k}) & t_{23}\epsilon_{\perp}({\bf k}) & 0 \\
0 & t_{32}\epsilon_{\perp}({\bf k}) & t_{33}\epsilon_{\parallel}({\bf k}) & \cdots \\
0 & 0 & \cdots & \cdots
\end{array}
\right). \label{eq:epsilon}
\end{eqnarray}
The intralayer (parallel) hopping and the interlayer (perpendicular)
hopping are specified by $t_{\alpha\alpha}\epsilon_{\parallel}({\bf k})$ and
$t_{\alpha\beta}\epsilon_{\perp}({\bf k})$, respectively,\cite{hoppings} with
\begin{equation}
\epsilon_{\parallel}=-2[\cos (k_x)+\cos (k_y)], \,\,\,\, \vert \epsilon_{\perp}({\bf {k}})\vert^{2}=1.
\end{equation}
Enforcing the separation between the surface ($S$) layers and the substrate ($L$), we can write  \eref{eq:green1} in a block form:
\begin{equation}
\left(
\begin{array}{cc}
{\bm A}_{LL} & {\bm A}_{LS} \\
{\bm A}_{SL} & {\bm A}_{SS}
\end{array}
\right).
\left(
\begin{array}{cc}
{\bm G}_{LL} & {\bm G}_{LS} \\
{\bm G}_{SL} & {\bm G}_{SS}
\end{array}
\right)=
\left(
\begin{array}{cc}
{\bm 1} & 0 \\
0 & {\bm 1}
\end{array}
\right),
\label{eq:green2}
\end{equation}
It should be noted that ${\bm A}_{LS}^{\dagger} = {\bm A}_{SL}$ and ${\bm A}_{LS}$ is a sparse matrix independent of $\bf k$ and $\omega_n$. For nearest-neighbor hopping, ${\bm A}_{LS}$ has only one nonzero entry corresponding to the hopping between the lowest layer of the surface and the top of the substrate.
We can solve \eref{eq:green2} to obtain the surface Green's function ${\bf G}_{SS}$. One obtains the following relation between $N\times N$ matrices defined on the surface block:
\begin{equation}
\left({\bm A}_{SS}- {\bm A}_{SL}{\bm A}_{LL}^{-1}{\bm A}_{LS}\right){\bf G}_{SS}= {\bm 1},
\label{eq:green3}
\end{equation}
The second term in the parenthesis of \eref{eq:green3} defines the embedding potential due to coupling of the surface region to the substrate. By definition ${\bm A}_{LL}^{-1}$ is the Green's function of the substrate decoupled from the surface
\begin{equation}
{\bf G}({\bf k}, i\omega_n) = \big[(i\omega_n+\mu){\bm 1}-{\bm
\epsilon}({\bf k})-{\bf \Sigma}(i\omega_n) \big]^{-1}.
\label{eq:green}
\end{equation}
The embedding potential then reads
\begin{equation}
{\bf S}({\bf k}, i\omega_n) = {\bm A}_{SL}{\bf G}({\bf k}, i\omega_n){\bm A}_{LS}.
\label{eq:embedding}
\end{equation}
Since $\bm {A}_{LS}$ is nonzero only between nearest-neighbor layers
of the substrate and surface regions, only the Green's function of the
first layer of the substrate,\cite{Kalkstein} i.e., the first entry ${\bf G}_{11}({\bf k}, i\omega_n)$ of \eref{eq:green},
is needed to calculate the embedding potential. ${\bf G}_{11}({\bf k}, i\omega_n)$ is computable directly using a recursive relation.\cite{Kalkstein} The self-energy appearing in \eref{eq:green} is obtained through a standard DMFT calculation for the bulk crystal corresponding to the substrate.

After constructing the embedding potential of the substrate, ${\bf
S}({\bf k}, i\omega_n)$, we can compute the self-energy of the surface
layers by DMFT. This can be achieved via the following steps: (i)
we associating an effective impurity model with each layer in the
surface region, and solve them by using an impurity solver to find the
layer-dependent local self-energies, $\Sigma_{\alpha}(i\omega_n)$. Then we
construct the surface region self-energy matrix which is
diagonal in layer indices $(\alpha, \beta)$ with the elements,
$\Sigma_{\alpha
\beta}(i\omega_n)=\Sigma_{\alpha}(i\omega_n)\delta_{\alpha \beta}$,
(ii) we calculate the on-site layer-dependent Green's function via
the following relation:
\begin{multline}
G_{\alpha}(i\omega_n)=\\ \sum_{{\bf k}}\left(
\frac{1}{(i\omega_n+\mu) {\bm 1}-{\bm \epsilon}({\bf k})-{\bf
S}({\bf k}, i\omega_n)-{\bm \Sigma}(i\omega_n)} \right)_{\alpha
\alpha}, \label{eq:G}
\end{multline}
where the $N\times N$ ${\bm \epsilon}({\bf k})$ matrix is given by
\eref{eq:epsilon}. (iii) We implement the DMFT self-consistency
relation for each layer, ${\mathcal G}^0_{\alpha}( i\omega_n) =
\big[G_{\alpha}^{-1}(i\omega_n) +
\Sigma_{\alpha}(i\omega_n)\big]^{-1}$, which determines the bath
parameters for the new  effective impurity model. These steps have to
be repeated until self-consistency is achieved.

The embedding method requires that we consider a relatively small number of surface layers; it is therefore a computationally less expensive extension of DMFT in the presence of an interface compared to the slab method, in which the inhomogeneous system is simply represented as a finite number of layers.
In this study, the number of surface layers is chosen to be $N=5$ and we tested (by varying this number) that this number provides converged  results.
\bbf{Our impurity solver is exact diagonalization,\cite{Caffarel} where the bath is
represented in terms of a finite number of levels, $n_s$. For the case
of phonon degrees of freedom we considered here, the infinite phonon
space is also truncated allowing for a maximum number of excited
phonons $n_{ph}$. The typical values we considered for the bath level
are $n_s=8$ and typical maximum number of phonons are
$n_{ph}=30-50$. We tested that those numbers provide essentially
converged results. For example changing $n_s$ from $8$ to $9$ changes $z$ only by $4\%$ for $g=0.5$ which is close to the transition. For smaller $g$ the error is smaller.}

\section{Results}
We use the technique explained in the previous section to study the Holstein model in a semi-infinite bipartite simple cubic lattice with in-plane translational symmetry and layer-dependent Hamiltonian parameters.
We will work at half-filling (one electron per site), where any charge
modulation is excluded by the particle-hole
symmetry\cite{charge_transfer} and local occupations on any layer,
including the surface, coincide with the average filling, $\langle
n_{\alpha}\rangle=1$. \bbf{We set the phonon frequency $\Omega_0 = 0.2t$,
which puts the system in the adiabatic regime.}
In order to characterize the metal-insulator transition, we use the quasiparticle weight, $z_{\alpha}=\left[ 1-\partial\Sigma_{\alpha}(\omega)/\partial\omega\vert_{\omega=0}\right]^{-1}$ [$\Sigma_{\alpha}(\omega)$ is the self-energy for layer $\alpha$] whose vanishing marks the transition to the insulating state in which there is no spectral weight at the Fermi level.
Another important quantity is the double occupancy, $d_{\alpha}=\langle n_{\alpha\uparrow}n_{\alpha\downarrow}\rangle$,
which is large in bipolaronic states.

We can model the inhomogeneity of the system and the different properties of the surface layer
by introducing layer-dependent parameters. \bbf{In particular, we can introduce different intralayer hopping $t_{11}$ or electron-phonon
coupling $g_1$ at the surface or we can tune the hopping between the surface and the second layer $t_{12}$.}

\bbf{One immediately realizes that the actual behavior of the parameters at
the surface will depend on the specific properties of each material
and on the geometry of the interface. On the other hand the aim of this work
is to understand general tendencies of an electron-phonon system in
the presence of a surface. Namely, we want to understand what happens
when the surface is more metallic than the bulk and when the opposite
situation is realized. 
Therefore we will use one single parameter, $t_{11}/t$, to model the effect of all the others. The case $t_{11} < t$
will represent all the situations in which the surface is less metallic than the bulk, while $t_{11} > t$ will
represent the opposite situation of a more metallic surface.}

\begin{figure}
\begin{center}
\center{\includegraphics[width=\normwidth]{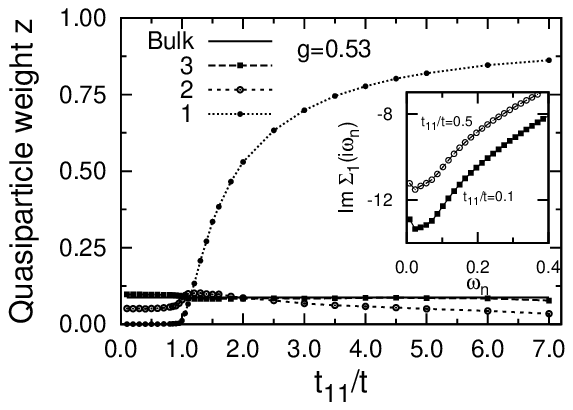}}\\
\center{\includegraphics[width=\normwidth]{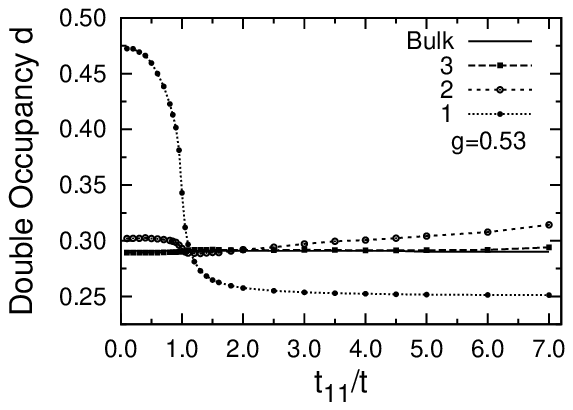}}
\caption{Layer-dependent quasiparticle weight $z_{\alpha}$ (top panel)
  and  layer-dependent double occupancy $d_{\alpha}$ (bottom panel) as
  a function of modified intra-layer surface hopping $t_{11}$. We show
  results for the first three layers of the semi-infinite Holstein
  model with simple cubic (001) surface geometry and for the bulk
  quasiparticle weight. $\alpha=1$ indicates the topmost surface
  layer.  The solid lines show bulk calculations. 
The insets show the imaginary part of the surface self-energy $Im\Sigma_{1} (i\omega_n)$ 
on the discrete mesh of the imaginary energies   $\omega_n=(2n+1)\pi/\tilde{\beta}$ ($\tilde{\beta}=400$)
for $t_{11}/t=0.1$ and $t_{11}/t=0.5$. $Im \Sigma_{\alpha=1}(i\omega_n)$ shows an upturn at small frequencies, compatible with a Fermi-liquid behavior.
}

\label{fig:fig1} 

\end{center}
\end{figure}

In \fref{fig:fig1} we show the evolution of the layer-dependent quasiparticle weights, $z_{\alpha}$ as a function
of the ratio $t_{11}/t$  for an e-ph coupling slightly smaller than the bulk critical coupling for the bipolaronic metal-insulator transition, $g_{c,bulk}\approx 0.55$.

We first focus on the regime in which the surface is less metallic than the bulk, i.e., $t_{11} < t$. We obviously
find that the surface quasiparticle weight $z_{\alpha=1}$ is smaller than that of the inner layers, although it 
does not vanish even for $t_{11}=0$, even though the e-ph coupling is very close to the bulk critical coupling. 
The finite (even if very small) value of $z_{\alpha=1}$ can be better identified by inspection of the 
self-energy which has to diverge to have a vanishing quasiparticle weight. It is seen from the inset of 
\fref{fig:fig1} that the surface self-energy $\Sigma_{\alpha=1}
(i\omega_n)$ goes up at low frequency and we expect extrapolation to
zero as $\omega_n \to 0$ in a Fermi liquid manner while a
divergence is excluded.

This behavior is not unexpected because a  metallic bulk is indeed able to determine an exponentially damped quasiparticle 
weight in the neighboring layers including the surface layer. This rules out the possibility to observe a truly insulating
surface on top of a metallic bulk, even if polaronic effects will be amplified on the surface.\cite{Nourafkan}
The double occupancy, shown in the bottom panel of \fref{fig:fig1}, presents a strong enhancement at the surface layer with respect to all the other layers and the bulk, as expected by the reduced hopping which favors the e-ph coupling. The second and third layers present only small deviations with respect to the bulk.

We now consider the case of a surface which is more metallic than the bulk, either because the surface e-ph coupling is
smaller or as we now analyze, $t_{11}>t$.
The results, also reported in \fref{fig:fig1}, show that, for very large $t_{11}$  $z_{\alpha=1}$ approaches
the free-electron value $z_{\alpha=1}=1$. This signals that the surface layer is essentially
decoupled from the rest of the system and it supports an uncorrelated motion of the electrons.
The rest of the system, however, remains strongly interacting and the $\alpha=2$ layer
represents the new surface layer, the $\alpha=3$ layer becomes the first subsurface layer and so on.
As is shown in \fref{fig:fig1}, for all values of $t_{11}/t$, the dependence
of the quasiparticle weight in the subsurface layers on $t_{11}$ is comparatively weak and
quickly diminishes with increasing distance from the surface.
The behavior of the double occupancy confirms the decoupling of the topmost layer, which, for
large values of $t_{11}/t$, approaches the noninteracting value, $0.25$.

For $g<g_{c,bulk}$, the same qualitative behavior is observed by changing the  
inter-layer surface hopping $t_{12} \neq t$ or by changing the e-ph coupling at the surface $g_{1} \neq g$.

\begin{figure}
\begin{center}
\includegraphics[width=\normwidth]{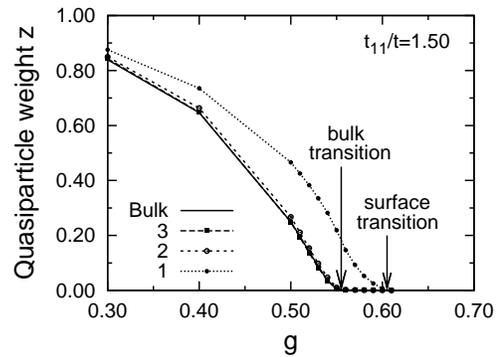}
\caption{$g$ dependence of quasiparticle weights $z_{\alpha}$ of
semi-infinite Holstein model for simple cubic lattice in the (001)
orientation for enhanced intra-layer surface hopping. Surface
transition at $g=g_{c,surface}$. Bulk transition at
$g=g_{c,bulk}$. } \label{fig:fig2}
\end{center}
\end{figure}

The decoupling between the surface and the bulk for large $t_{11}$ implies that 
one can in principle approach the situation where a metallic surface
coexists with an insulating bulk. To investigate this possibility, we computed
the quasiparticle weights as a function of $g$. In \fref{fig:fig2} we plot
the quasiparticle weight vs. $g$ for a moderately enhanced surface hopping rate, $t_{11}=1.5t$.
Upon increasing $g$ two different critical interactions are found. The first one marks the transition from a metallic to a bipolaronic insulating state at $g_c=g_{c,bulk}$, in which all the bulk quasiparticle weights (all layers except the surface) vanish. For larger e-ph interaction there is a range of values of $g$ in which the bulk is a bipolaronic insulator while the surface is still metallic with a finite $z_{\alpha=1}$. Indeed, in this region some weight is induced in the subsurface layers. Since the low energy surface excitations cannot propagate into the bulk for $g>g_{c,bulk}$ and are instead reflected back to the surface for energies below the bulk excitation gap, the induced quasiparticle weight decreases exponentially with increasing distance from the surface.

At a second critical coupling, $g_{c,surface}$, the surface also becomes insulating and bipolaronic. For 
$g>g_{c, surface}$ the entire system is in the bipolaronic insulating phase. A rather moderate enhancement of $t_{11}$ is sufficient to obtain a metallic surface phase. Obviously, a larger $t_{11}$ means that electrons in the first layer are more itinerant. A smaller surface coordination number clearly counteracts this mechanism. Consequently, we expect that a larger $t_{11}$ is needed to obtain a metallic surface state for more open surfaces, such as, for example, the (110) surface.
The range of coupling $g$ where a metallic surface coexists with an insulating
bulk quickly increases as $t_{11}$ is increased. For $t_{11}
\to \infty$  the bulk energy scales become irrelevant and the
electronic structure of the surface layer decouples from the rest of
the system.

\bbf{The overall results obtained here for a model with electron-phonon
interactions are qualitatively similar to those of
Ref. \onlinecite{Potthoff2} for a repulsive Hubbard model despite the fact that the nature
of the transition is different in the two models. In both cases one can have an
insulating surface coexisting with an insulating bulk, which is a Mott
insulator in the Hubbard model and a bipolaronic insulator in the
Holstein model. Instead, in both models a more insulating surface gives rise to
a single metal-insulator transition. Indeed the similarity
between the two cases is not accidental.  In the antiadiabatic limit
the Holstein model becomes the attractive Hubbard model, which,
at half-filling, can be mapped onto a repulsive model by a
particle-hole transformation. Then the Mott transition of the
repulsive model becomes a pairing transition in which fermionic pairs
are formed,\cite{pairingmit} which in turn corresponds to the bipolaronic transition in the Holstein model. 
Therefore the main difference between the two models is the retarded nature of the electron-phonon interaction, as opposed to the instantaneous
Hubbard interaction. 
In principle in the electron-phonon case, larger lattice distortions at the surface\cite{Nourafkan} could favor an
insulating surface with respect to the case of the Hubbard model.
However our results show that the dynamical nature of the
interaction is not able to introduce qualitative differences with
respect to a purely electronic model.}

\begin{figure}
\begin{center}
\center{\includegraphics[width=\normwidth]{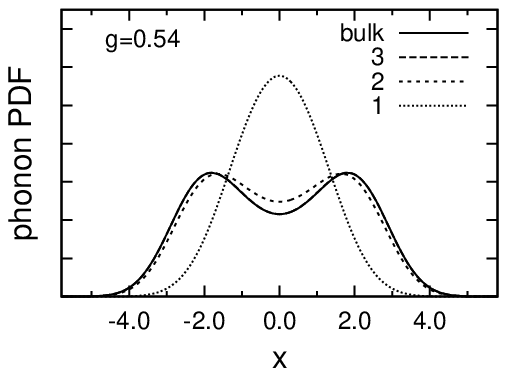}}\\
\center{\includegraphics[width=\normwidth]{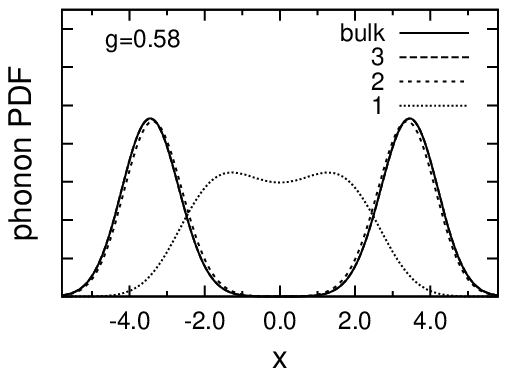}}
\caption{Phonon probability-distribution function for the first three layers of the semi-infinite Holstein model with nonuniform model parameters and the bulk phonon probability-distribution function for two different values of e-ph coupling strength, $g$. We used $t_{11}=1.5t$.} \label{fig:fig3}
\end{center}
\end{figure}

To gain further insight about the region in which the surface remains metallic, we consider the behavior of the phonon displacement probability distribution function (PDF), $P(x)=\langle \phi_0|x\rangle\langle x|\phi_0\rangle$, where $|\phi_0\rangle$ is the ground-state wave function and $|x\rangle\langle x|$ is the projection operator on the subspace where the phonon displacement at a given site $\hat{x}$ has value $x$.

This quantity is a measure of the distribution of the local distortions.\cite{PDFCalculation} In the absence of e-ph interaction, $P(x)$ is a Gaussian centered around $x=0$. A small e-ph coupling slightly broadens the distribution which remains centered around $x=0$, implying that the coupling is not sufficient to give rise to a finite polarization of the lattice. Continuously increasing the interaction one eventually obtains a bimodal distribution with two maxima at $x = \pm x_0$. A bimodal PDF indicates that a certain number of the lattice sites are polarized by the presence of electrons in such a way that the average value of the polarization is greater than its fluctuations and therefore provide evidence for electron/phonon entanglement, i.e., a polaronic state. The point at which the phonon PDF becomes bimodal is used as a marker of the polaron crossover,\cite{Ciuchimanga} while $P(x=0)=0$ can be used to characterize the transition to the bipolaronic insulator (even if the vanishing of $z$ is a more rigorous criterion).
Figure \ref{fig:fig3} shows the phonon PDF for $t_{11}=1.5t$ at two e-ph coupling values, one for $g<g_{c,bulk}$   (top panel) and one for $g_{c,bulk}<g<g_{c,surface}$ (lower panel). In the first case the system is metallic but polarons are already formed in all the layers except the topmost one, which has a larger hopping amplitude, $t_{11}$. Notice that the appearance of polaronic distortion is not sufficient to make the bulk insulating, confirming that the polaron crossover and the bipolaronic transition do not coincide. Upon increasing the e-ph coupling and for  $g_{c,bulk}<g<g_{c,surface}$, the phonon PDF of all layers except the surface go to zero at $x=0$ (bottom panel). This confirms the insulating phase of these layers in this range of couplings while the surface layer shows a metallic state with polaronic character, as shown by the only slightly bimodal PDF.

\section{Concluding Remarks}
We have investigated the effect of a surface on a strongly coupled electron-phonon system. We described this system with a Holstein model on a cubic lattice cut along one of the coordinate axes, assuming that the surface parameters are different from the bulk ones. We used the ratio between the hopping within the surface layer and the hopping within bulk layers, $t_{11}/t$, to represent the effects of other nonuniform parameters and considered both situations in which the surface is less metallic than the bulk ($t_{11} < t$) and the opposite regime of a more metallic surface.
Our focus is on the strong coupling regime, where we ask whether bulk
and surface can be decoupled as far as the transport properties are
concerned. In particular, for a given set of parameters, we ask
whether one part of the system can be insulating while the other is
metallic. \bbf{We considered the system at half-filling, where the
electron-phonon interaction can drive a bipolaronic phase
transition. This choice inhibits charge transfer between the surface
and the bulk.}
Our investigation, based on the embedding method for DMFT,\cite{Ishida} shows that a metallic surface can coexist with an insulating bulk when $t_{11}/t > 1$ already for moderate values of this ratio while the opposite behavior is not realized even when the surface hopping is vanishing. The bulk excitations are always able to penetrate in the surface layer, even if they are strongly damped.
Polaronic distortions, measured by the phonon distribution function, can be significantly different in the surface over a range of parameters.

\bbf{From the experimental point of view, even if the most typical
situation is that in which the surface is more insulating than the
bulk, evidence for a ferromagnetic metallic surface has been
reported in antiferromagnetic insulating manganites\cite{dagotto}  and
a surface insulator-to-metal transition has been observed in insulating
NiS$_2$.\cite{ddsarma} In general these measurements are difficult and rely
on an interpretation of transport and magnetic measurements on samples with
varying grain sizes.\cite{biswas2005} Similarly, a use of photoemission and tunneling
spectroscopies should be helpful to discern surface vs bulk properties.} 

\begin{acknowledgments}
This work was supported in part by the Natural Sciences and Engineering
Research Council of Canada (NSERC), by ICORE (Alberta), and by the Canadian
Institute for Advanced Research (CIfAR).
M.C.'s activity is funded by the European Research Council under FP7/ERC Starting Independent Research Grant ``SUPERBAD'' ( Agreement No. 240524) and MIUR PRIN 2007 under Grant No. 2007FW3MJX003.

\end{acknowledgments}

\bibliographystyle{prsty}

\begin{thebibliography}{99}
\bibitem{hoppings} From now on we drop the site index from the hopping matrix elements which are only labeled by the layer index.
\bibitem{charge_transfer}
For a nonbipartite (e.g., fcc) lattice a charge modulation is possible also at half-filling.
\bibitem{PDFCalculation}
Using exact diagonalization as an impurity solver we can compute the PDF as $P(x)=\sum_{nm}\psi_n (x) \psi_m (x) \langle \phi_0|n\rangle \langle m|\phi_0\rangle$, where $\vert\phi_0\rangle$ is the impurity model ground state vector, $|n\rangle$ and $|m\rangle$ are the eigenstates of the harmonic oscillator and $\psi_n (x)$ are the corresponding eigenfunctions.
\end{thebibliography}


\end{document}